\documentclass[aps,twocolumn,epsfig]{revtex4}
\usepackage{amsfonts}
\usepackage{epsfig}
\usepackage{color}
\usepackage{graphicx}
\usepackage{dcolumn}
\usepackage{bm}

\begin{document}
\title{Rotational stabilization and destabilization of an optical cavity}
\author{Steven~J.~M.~Habraken and Gerard Nienhuis}
\affiliation{Leiden Institute of Physics, Leiden University, P.O.
Box 9504, 2300 RA Leiden, The Netherlands}

    \begin{abstract}
We investigate the effects of rotation about the axis of an
astigmatic two-mirror cavity on its optical properties. This
simple geometry is the first example of an optical system that can
be destabilized and, more surprisingly, stabilized by rotation. As
such, it has some similarity with both the Paul trap and the
gyroscope. We illustrate the effects of rotational
(de)stabilization of a cavity in terms of the spatial structure
and orbital angular momentum of its modes.
    \end{abstract}

\maketitle

Instability is ubiquitous in physics. Examples range from the
simple case of a particle on the top of a hill to complex weather
systems. Some partially unstable systems can be stabilized by
external motion. One of the best-known examples of dynamical
stabilization is the Paul trap \cite{Paul53}, which is similar to
rotational stabilization of a particle in a saddle-point potential
\cite{Thompson02}. Another well-known example of rotational
stabilization is the gyroscope. Similar behavior has also been
observed in thermodynamically large systems such as granular
matter \cite{Zik94} and fluids \cite{Salhi07}.

In recent years, optical cavities with moving elements have become
topical. State-of-the-art experiments focus on optomechanical
oscillators driven by radiation pressure
\cite{Kippenberg05,Corbitt06} and cavity-assisted trapping and
cooling \cite{Cohadon99,Kleckner06,Groblacher08}. Possible
applications range from weak-force detection \cite{Lucamarini06}
to fundamental research on quantum entanglement
\cite{Vitali07,Bhattacharya08} and decoherence
\cite{Bernad06,Kleckner08} on macroscopic scales. In addition to
the longitudinal radiation pressure, electromagnetic fields can
exert transverse forces on small particles due to their phase
structure \cite{Roichman08}. A specific example is the transfer of
optical orbital angular momentum \cite{Molina-Terriza07}, which
can can give rise to a torque along the propagation axis of the
beam. Recently, it was shown theoretically that this torque can be
sufficiently large to trap and cool the rotational degrees of
freedom of a mirror in a cavity-assisted setup
\cite{Bhattacharya07}. Here, we focus on the
complementary question: How does rotation of a mirror affect the
optical properties of a cavity and, in particular, its
(in)stability? As such, this work constitutes the first analysis
of rotational effects on stability in optics.

We consider a cavity that consists of two mirrors facing each
other. In the simplest case both mirrors are spherical. Depending
on their focussing properties, a ray that is coupled into such a
cavity can either be captured, or escape after a finite (and
typically small) number of round trips. In the latter case the
cavity is geometrically unstable whereas it is stable in the
first. The stability criterion for this system can be expressed as
\cite{Siegman}
\begin{equation}\label{stabcrit}
0<g_{1}g_{2}<1\;,
\end{equation}
where $g_{1,2}=1-L/R_{1,2}$ with $R_{1,2}$ the radii of curvature
of the two mirrors and $L$ their separation. The optical
properties of unstable cavities are essentially different from
those of their stable counterparts \cite{Siegman}. The modes of a
stable cavity are stationary and spatially confined whereas the 
'modes' of an unstable cavity are self-similar diverging patterns
that have a fractal nature \cite{Karman99}. Instability is a
necessary condition for an optical cavity to display chaotic
behavior \cite{Puentes04}.

\begin{figure}[!t]
\begin{center}
\includegraphics[width=4cm]{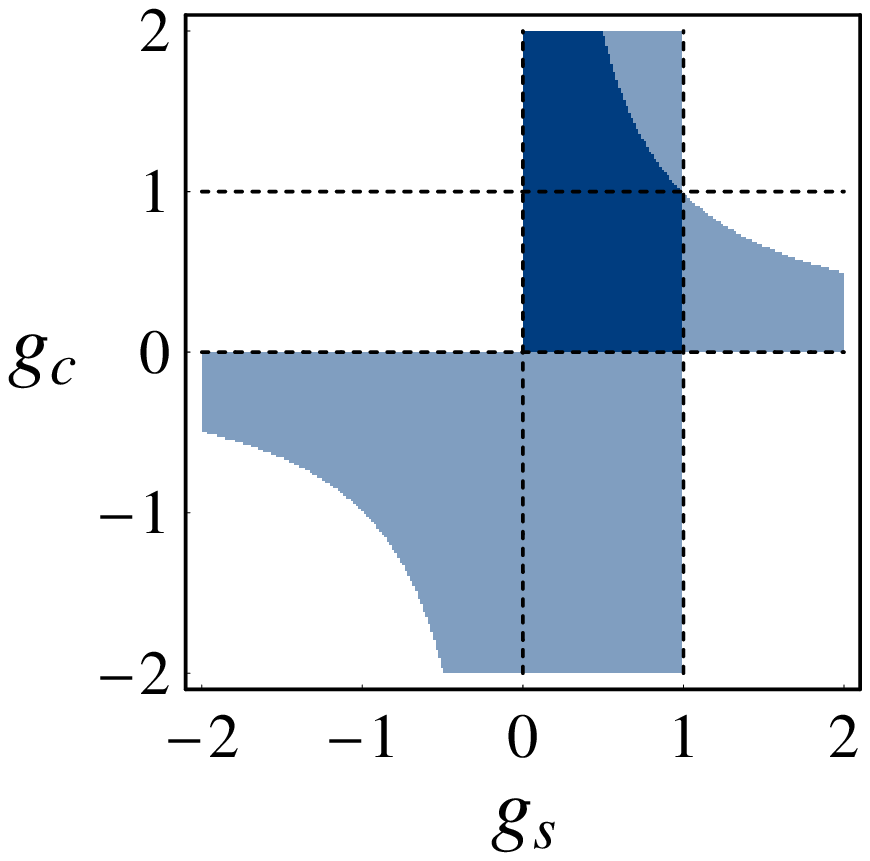}
\includegraphics[width=4cm]{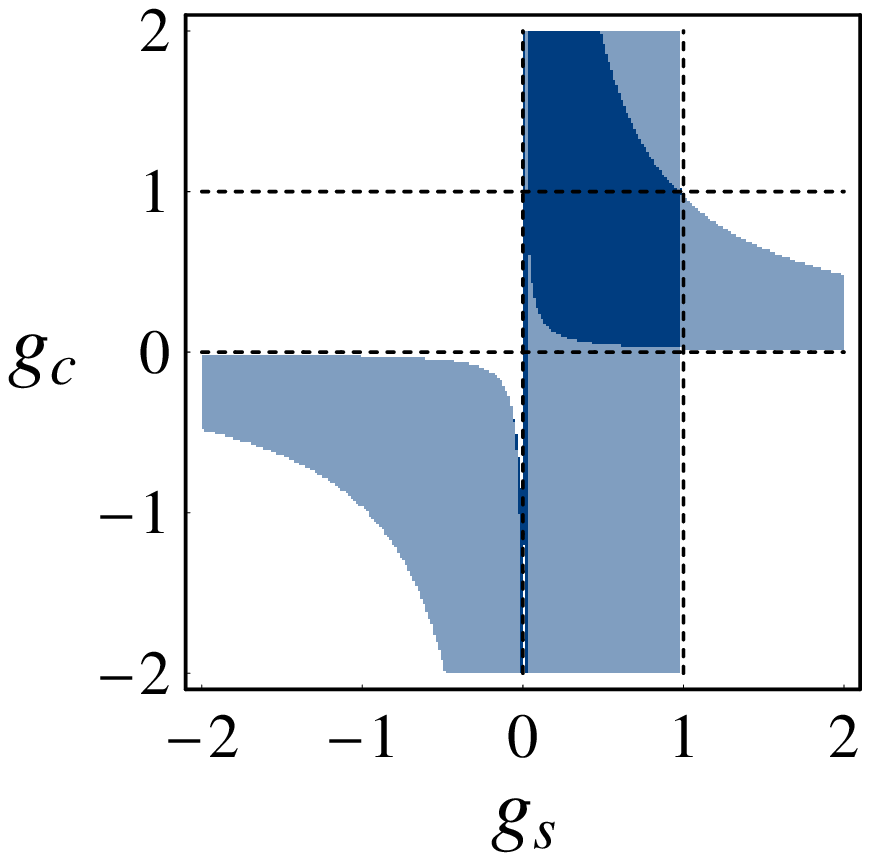}\\
\vspace{0.5cm}
\includegraphics[width=4cm]{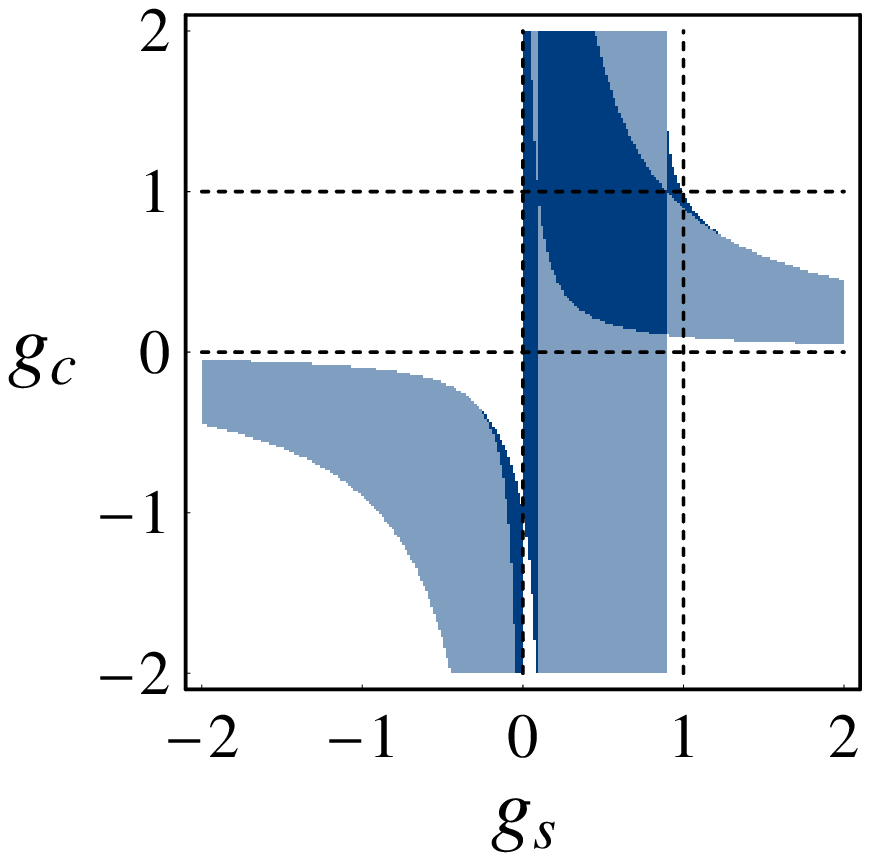}
\includegraphics[width=4cm]{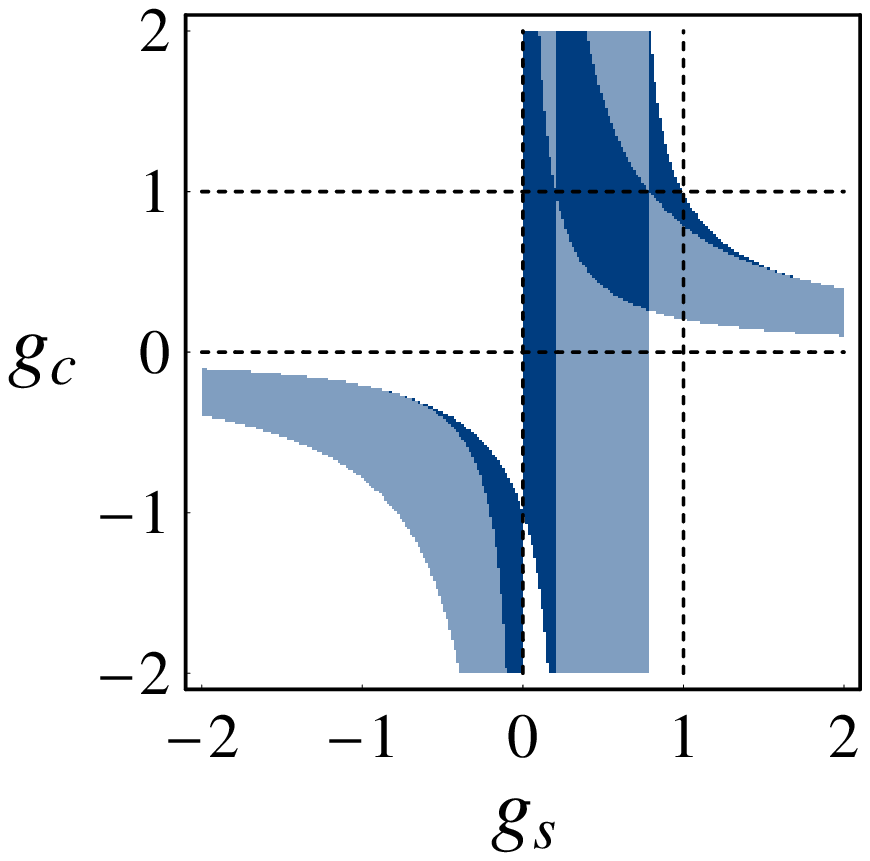}\\
\vspace{0.5cm}
\includegraphics[width=4cm]{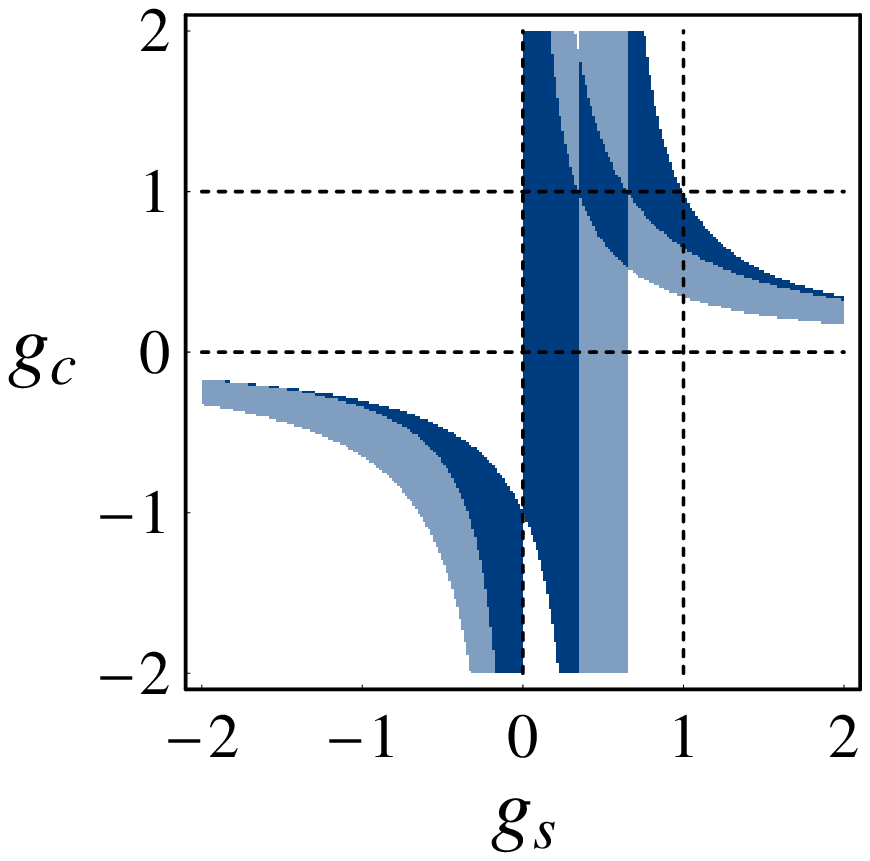}
\includegraphics[width=4cm]{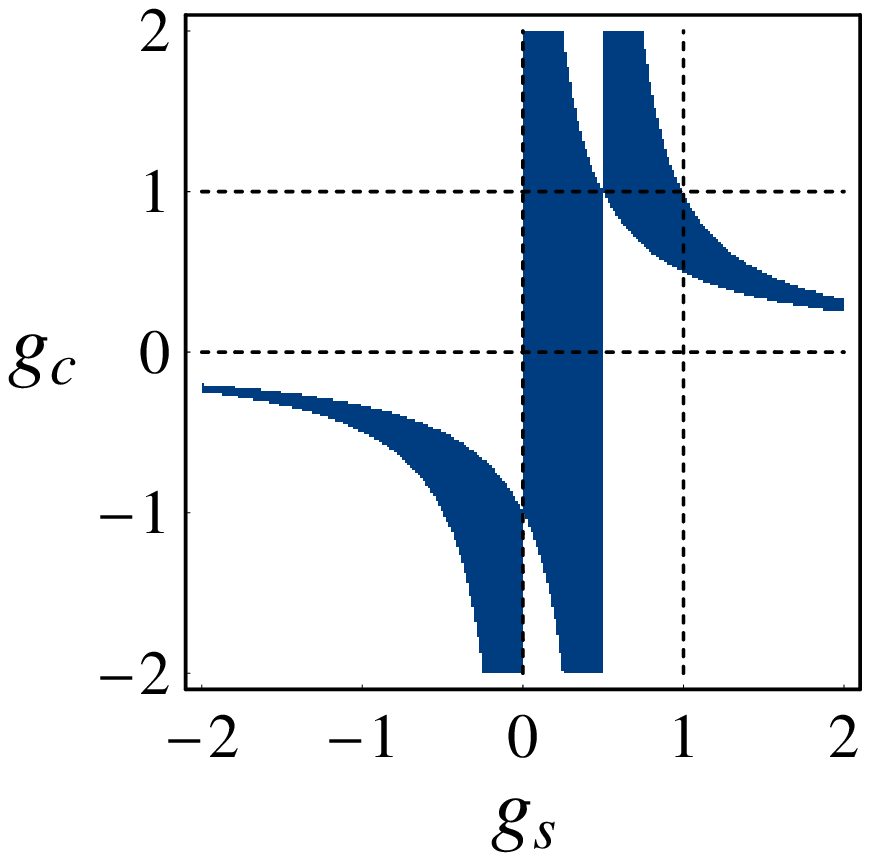}
\end{center}
\caption{\label{stab} Stable (dark), partially stable
(light) and unstable (white) areas of the configuration space
$(g_{\mathrm{s}},g_{\mathrm{c}})$ for a cavity that consists of a stationary spherical
and a rotating cylindrical mirror, for different rotation frequencies.
From left to right and from top to bottom the rotation
frequency is increased in equal steps $\Omega_{0}/20$ from $0$ to
$\Omega_{0}/4$.}
\end{figure}

We consider rotations about the optical axis of a cavity and
expect an effect only if at least one of the mirrors is astigmatic
(or cylindrical), so that the cavity lacks axial symmetry. In
general, both mirrors can be astigmatic with non-parallel axes
but, for simplicity, we focus on a cavity that consists of a
cylindrical ($\mathrm{c}$) and a spherical ($\mathrm{s}$) mirror.
The curvature of each mirror can be specified by a single
$g$ parameter so that the configuration space, spanned by
$g_{\mathrm{s}}$ and $g_{\mathrm{c}}$, is two dimensional. In the
absence of rotation the stability criterion in the plane through
the optical axis in which the cylindrical mirror is curved is of
the form of Eq. (\ref{stabcrit}):
$0<g_{\mathrm{s}}g_{\mathrm{c}}<1$. In the other, perpendicular
plane through the cavity axis, in which the cylindrical mirror is
flat, the stability criterion reads: $0<g_{\mathrm{s}}<1$. As is
indicated in the upper left plot of Fig. \ref{stab}, stable (dark)
areas appear where both criteria are met. The cavity is partially
stable (light) in areas where only one of the two is fulfilled.
When the cavity is partially stable, both a ray that is coupled
into it and its modes are confined in one of the two transverse
directions only. One may guess that rotation disturbs the
confinement of the light by the mirrors so that all (partially)
stable cavities will eventually lose stability if the rotation
frequency is sufficiently increased. However, we will show that
this is not the case.

In order to describe the diffraction of the light inside the
rotating cavity, we use the paraxial approximation and its
generalization to the time-dependent case \cite{Lax75,Deutsch91}.
We write the transverse electric field of a propagating mode as
\begin{equation}\label{Efield}
\mathbf{E}(\mathbf{r},t)=\mathrm{Re}\left\{E_{0}\mathbf{e}
u(\mathbf{r},t)e^{ikz-i\omega t}\right\}\;,
\end{equation}
where $E_{0}$ is the amplitude of the field, $\mathbf{e}$ is the
polarization, $k$ is the wave number and $\omega=ck$ is the
optical frequency with $c$ the speed of light. The large-scale
spatial structure and slow temporal variations of the electric
field are characterized by the complex scalar profile
$u(\mathbf{r},t)$. In lowest order of the paraxial approximation
and under the assumption that the time dependence of the profile
is slow compared to the optical time scale, the electric field is
purely transverse and the profile $u(\mathbf{r},t)$ obeys the
time-dependent paraxial wave equation
\begin{equation}\label{tdpweu}
\left(\nabla_{\rho}^{2}+2ik\frac{\partial}{\partial
z}+\frac{2ik}{c}\frac{\partial}{\partial
t}\right)u(\mathbf{r},t)=0\;,
\end{equation}
with $\nabla^{2}_{\rho}=\partial^{2}/\partial
x^{2}+\partial^{2}/\partial y^{2}$. If we omit the derivative with
respect to time, this equation reduces to the standard paraxial
wave equation, which describes the diffraction of a freely
propagating stationary paraxial beam. The additional time
derivative accounts for the time dependence of the profile and
incorporates retardation between distant transverse planes.

The dynamics of light inside a cavity is governed by the boundary
condition that the electric field vanish on the mirror surfaces.
For a rotating cavity, this boundary condition is explicitly time
dependent. This time dependence vanishes in a corotating frame
where it is sufficient to consider time-independent propagating
modes $v(\mathbf{r})$. The transformation that connects
$v(\mathbf{r})$ and $u(\mathbf{r},t)$ takes the form
\begin{equation}\label{rotmode}
u(\mathbf{r},t)=\hat{U}_{\mathrm{rot}}(\Omega t)v(\mathbf{r})\;,
\end{equation}
where $\Omega$ is the rotation frequency and
$\hat{U}_{\mathrm{rot}}(\alpha)=\exp(-i\alpha\hat{L}_{z})$ is the
operator that rotates a scalar function over an angle $\alpha$
about the $z$ axis with $\hat{L}_{z}=-i(x\partial/\partial
y-y\partial/\partial x)$ the $z$ component of the orbital angular
momentum operator. Substitution of the rotating mode
(\ref{rotmode}) in the time-dependent wave equation (\ref{tdpweu})
gives
\begin{equation}\label{tdpwev}
\left(\nabla_{\rho}^{2}+2ik\frac{\partial}{\partial
z}+\frac{2\Omega k}{c}\hat{L}_{z}\right)v(\mathbf{r})=0\;
\end{equation}
for $v(\mathbf{r})$. The transformation to a rotating frame gives
rise to a Coriolis term, in analogy with particle mechanics. Since
$\nabla^{2}_{\rho}$ and $\hat{L}_{z}$ commute, the formal solution of
Eq. (\ref{tdpwev}) can be expressed as
\begin{equation}\label{rotprop}
v(\mathbf{\rho},z)=\hat{U}_{\mathrm{f}}(z)\hat{U}_{\mathrm{rot}}\left(-\frac{\Omega
z}{c}\right)v(\rho,0)\equiv\hat{U}(z)v(\rho,0)\;,
\end{equation}
where $\rho=(x,y)$ and
$\hat{U}_{\mathrm{f}}(z)=\exp\big(\frac{iz}{2k}\nabla_{\rho}^{2}\big)$
is the unitary operator that describes free propagation of a
paraxial beam in a stationary frame. The operator $\hat{U}(z)$ has
the significance of the propagator in the rotating frame. The
rotation operator arises from the Coriolis term in Eq.
(\ref{tdpwev}) and gives the propagating modes a twisted nature.

The transformation of paraxial modes under propagation and optical
elements can be expressed in terms of a ray ($ABCD$) matrix
\cite{Siegman}. The standard $2\times 2$ ray matrices that
describe optical elements with axial symmetry can be found in any
textbook on optics. The ray matrix of a composite system can be
constructed by multiplying the ray matrices that describe the
optical elements and the distances of free propagation between
them, in the proper order. Generalization to astigmatic optical
elements is straightforward and requires $4\times 4$ ray matrices
\cite{Siegman,Habraken07}. The ray matrix that describes
propagation in a rotating frame is, analogous to Eq.
(\ref{rotprop}), given by
$M(z)=M_{\mathrm{f}}(z)M_{\mathrm{rot}}(-\Omega z/c)$, where
$M_{\mathrm{f}}(z)$ is the $4\times 4$ ray matrix that describes
free propagation over a distance $z$ and
$M_{\mathrm{rot}}(\alpha)$ is the $4\times 4$ ray matrix that
rotates the position $\rho$ and propagation direction $\Theta$ of
a ray $r=(\rho,\Theta)$ over an angle $\alpha$ about the $z$ axis.
Starting at the entrance plane of the spherical mirror, the
time-independent ray matrix that describes a round trip through
the rotating cavity in the corotating frame is then
\begin{equation}\label{mrt}
M_{\mathrm{rt}}=M(L)\cdot M_{\mathrm{c}}\cdot M(L)\cdot
M_{\mathrm{s}}\;,
\end{equation}
where $L$ is the mirror separation and $M_{\mathrm{s}}$ and
$M_{\mathrm{c}}$ are the ray matrices for the spherical and the
cylindrical mirror. They are fully determined by the radii of
curvature and the orientation of the mirrors in the transverse
plane \cite{Habraken07}.

Typically, the round-trip ray matrix (\ref{mrt}) has four distinct
time-independent eigenvectors $\mu_{i}$ with corresponding
eigenvalues $\lambda_{i}$. In the rotating frame, any
time-dependent incident ray
$r_{0}(t)=\left(\rho(t),\Theta(t)\right)$ can be expanded as
$r_{0}(t)=\sum_{i}a_{i}(t)\mu_{i}$. After $n$ times bouncing back
and forth between the mirrors, the ray evolves into
$r_{n}(t+2nL/c)=\sum_{i}a_{i}(t)\lambda_{i}^{n}\mu_{i}$. The
possibly complex eigenvalues have the significance of the
magnification of the eigenvector after one round trip and it
follows that a cavity is stable only if all four eigenvalues have
absolute value 1. The eigenvalues of any physical ray matrix come
in pairs $\lambda$ and $\lambda^{-1}$ \cite{Habraken07} so that
deviations from $|\lambda|= 1$ appear in two of the four
eigenvalues at the same time. If only two eigenvalues have
absolute value 1, the cavity is partially stable. The eigenvalues
of the round-trip ray matrix (\ref{mrt}) do not depend on the
frame of reference, and it follows that the same is true for the
notion of stability.

A ray that is bounced back and forth inside the cavity hits a
mirror at time intervals $L/c$. Since a rotation over $\pi$ turns
an astigmatic mirror to an equivalent orientation, it follows that
the stability of a cavity is not affected by a change in the
rotating frequency $\Omega\rightarrow\Omega+p\Omega_{0}$ with
integer $p$ and $\Omega_{0}=c\pi/L$. In the present case, in which
one of the mirrors is spherical, a ray hits the cylindrical mirror
at time intervals $2L/c$ so that the eigenvalues $\lambda_{i}$ are
periodic with $\Omega_{0}/2$. Moreover, an astigmatic cavity is
not gyrotropic so that the eigenvalues do not depend on the sign
of $\Omega$. It follows that it is sufficient to only consider
rotation frequencies in the range $0<\Omega<\Omega_{0}/4$.

\begin{figure}[!t]
\begin{center}
\includegraphics[width=1.8cm]{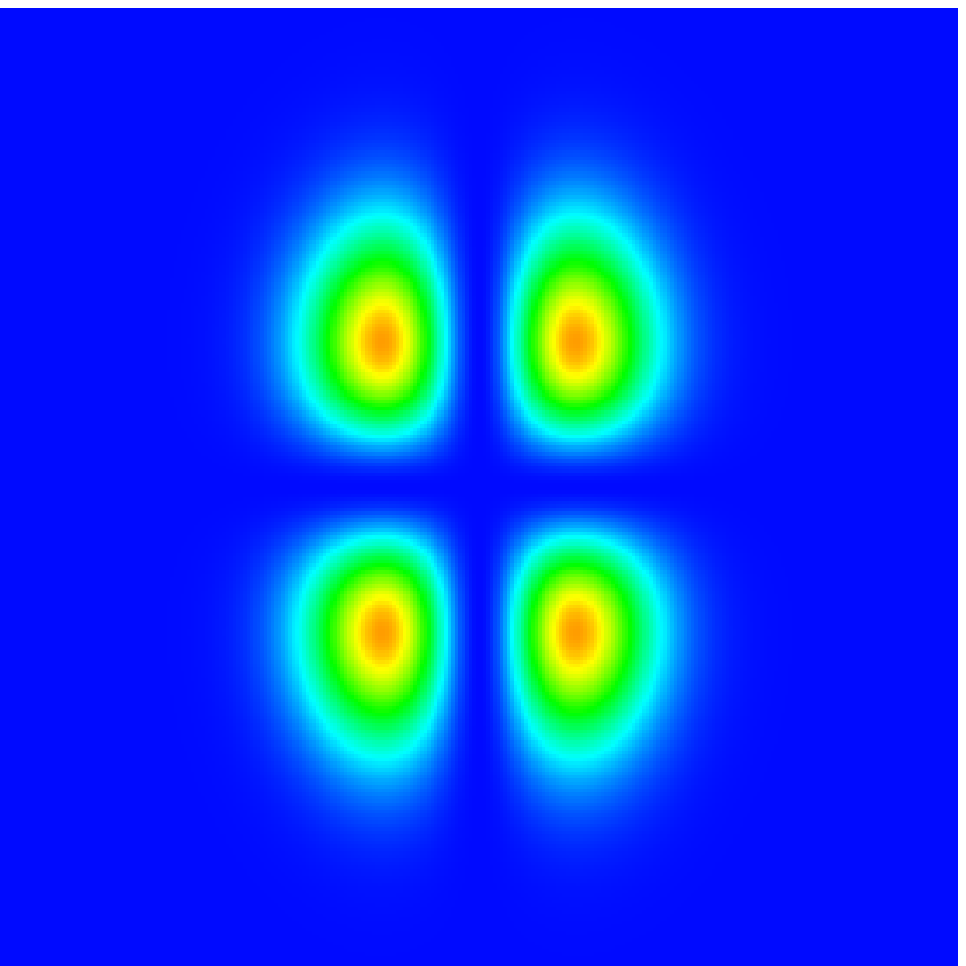}\hspace{0.2cm}
\includegraphics[width=1.8cm]{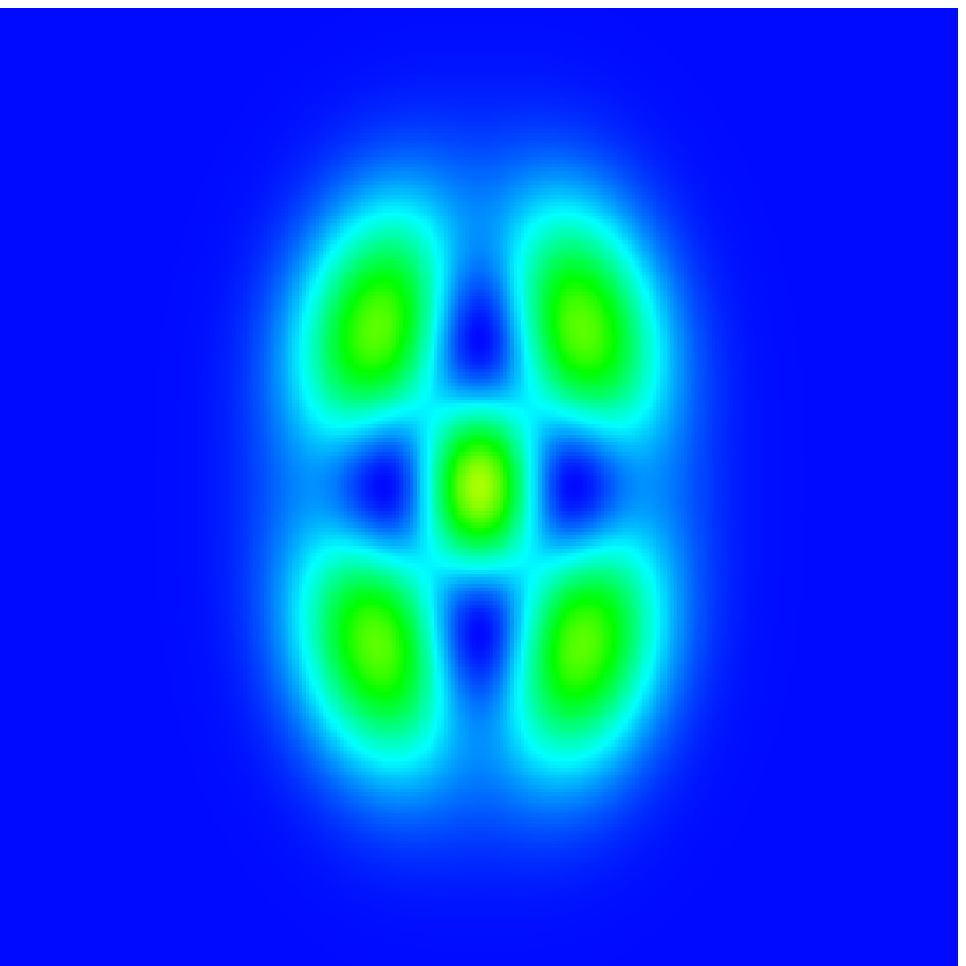}\hspace{0.2cm}
\includegraphics[width=1.8cm]{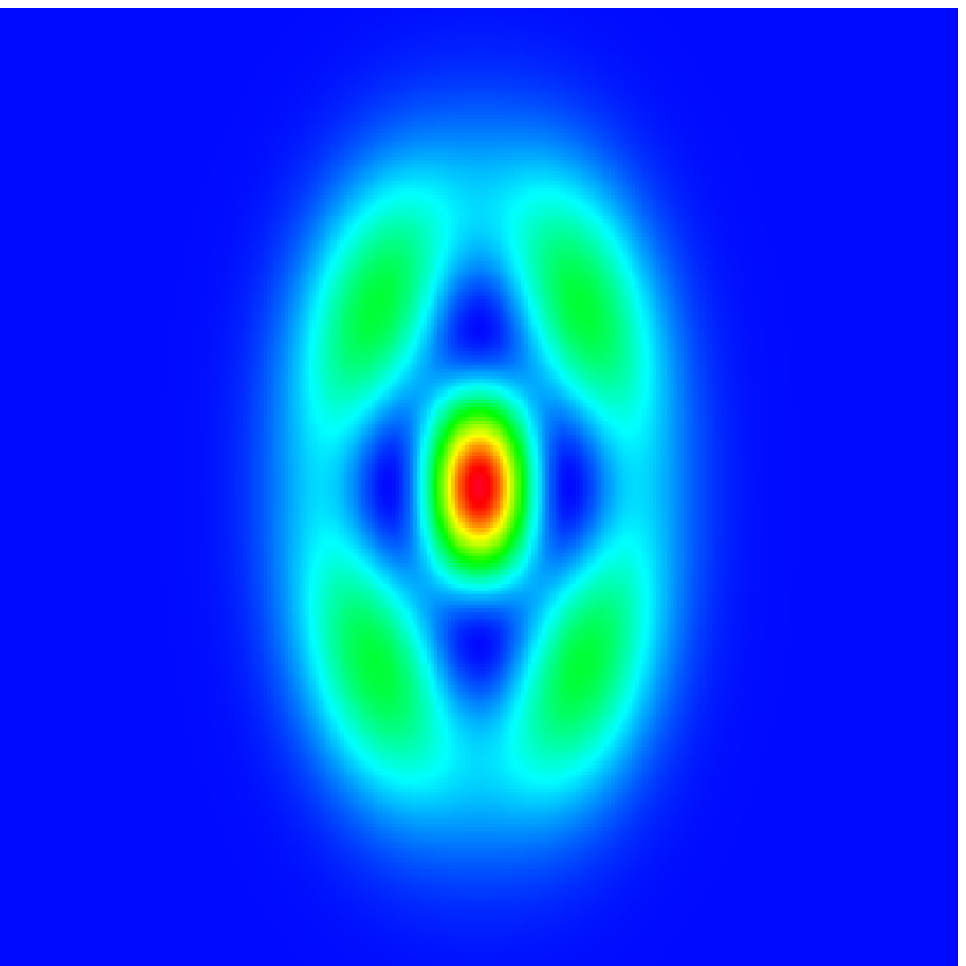}\hspace{0.2cm}
\includegraphics[width=1.8cm]{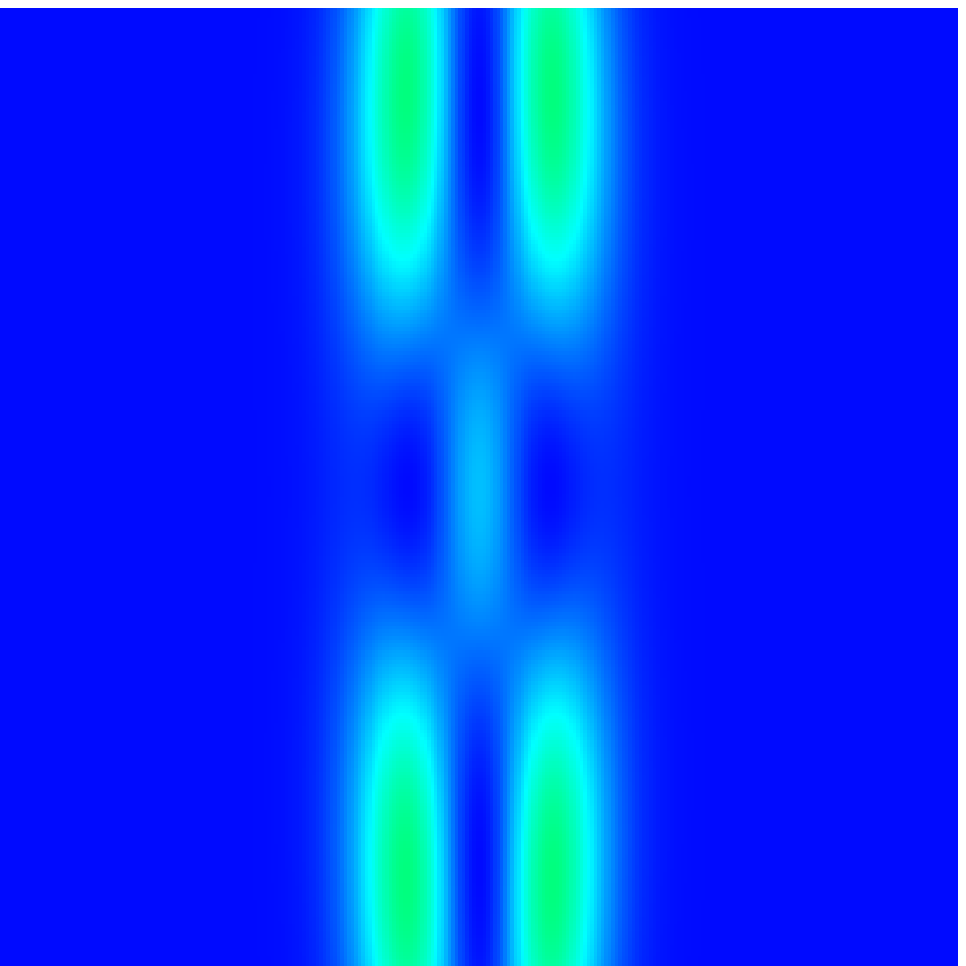}\\
\vspace{0.3cm}
\includegraphics[width=1.8cm]{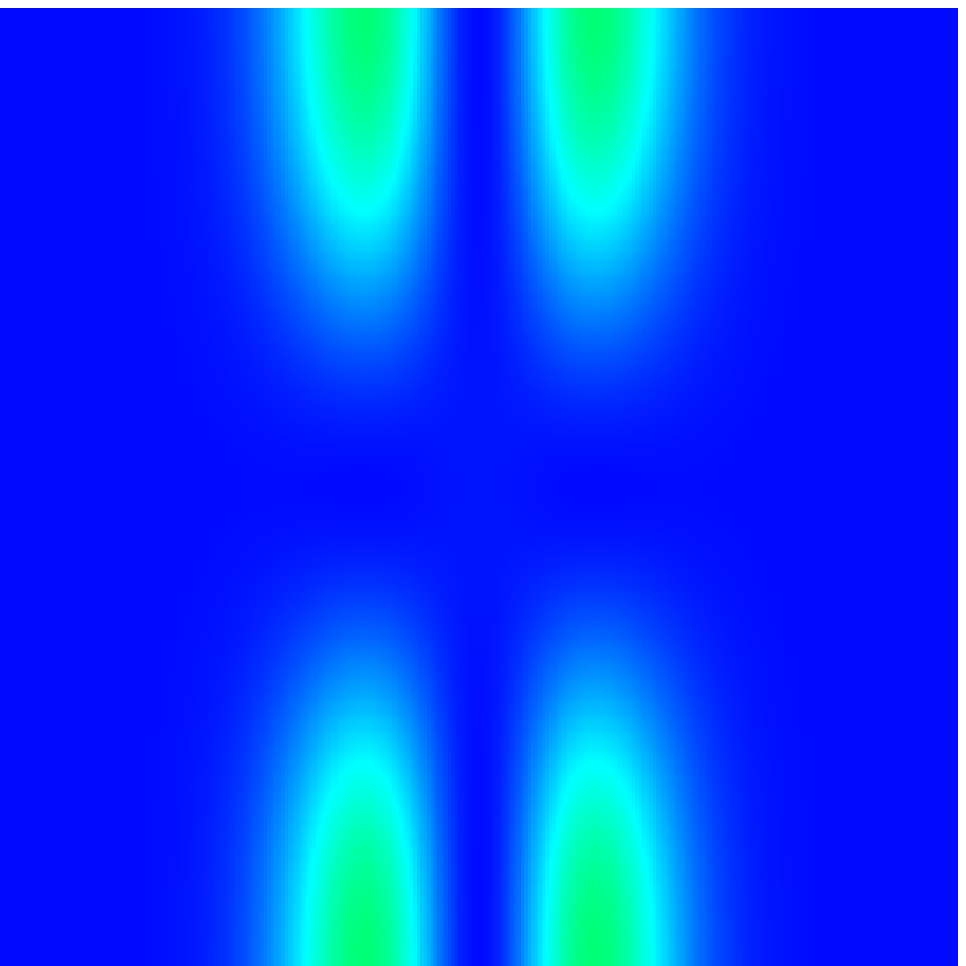}\hspace{0.2cm}
\includegraphics[width=1.8cm]{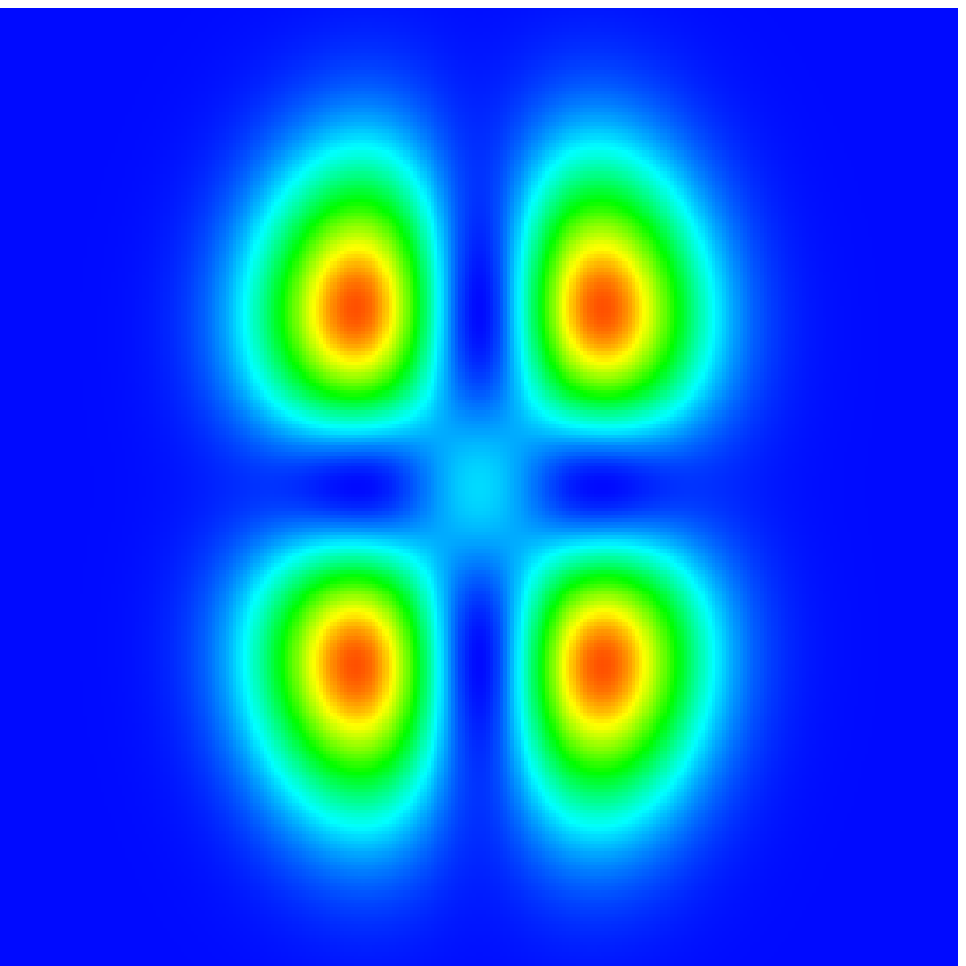}\hspace{0.2cm}
\includegraphics[width=1.8cm]{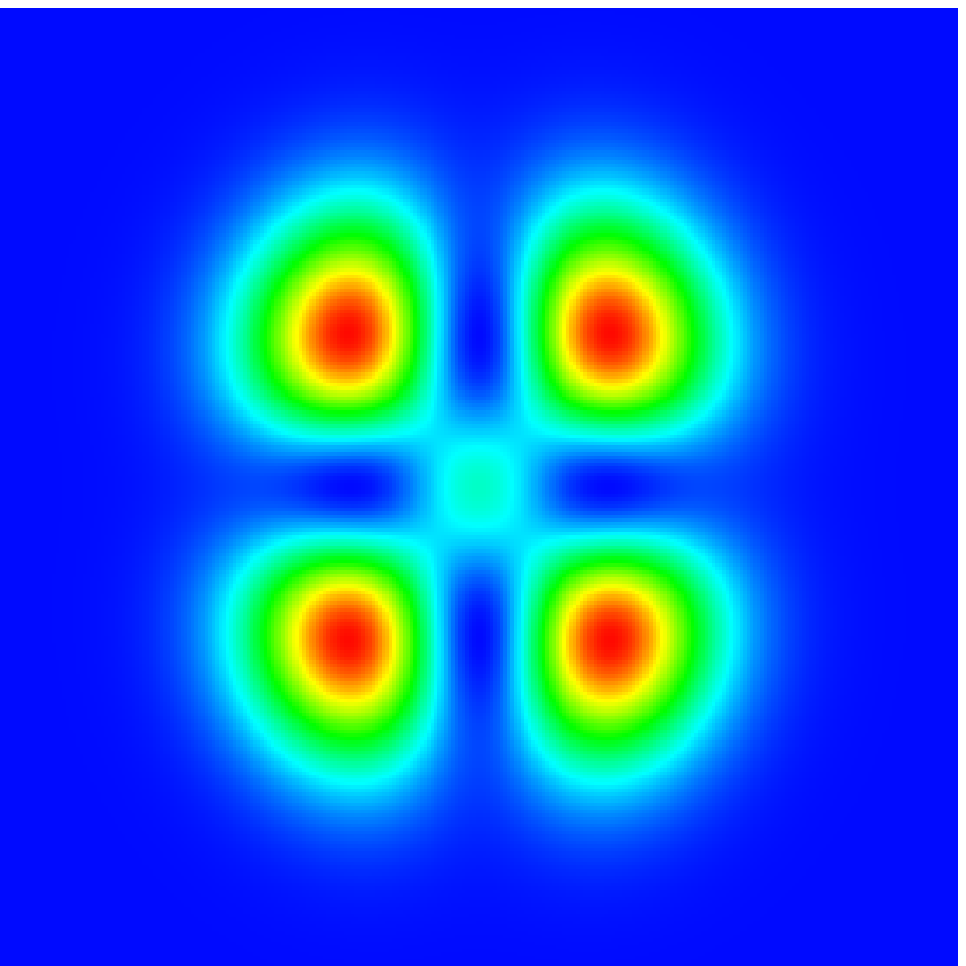}\hspace{0.2cm}
\includegraphics[width=1.8cm]{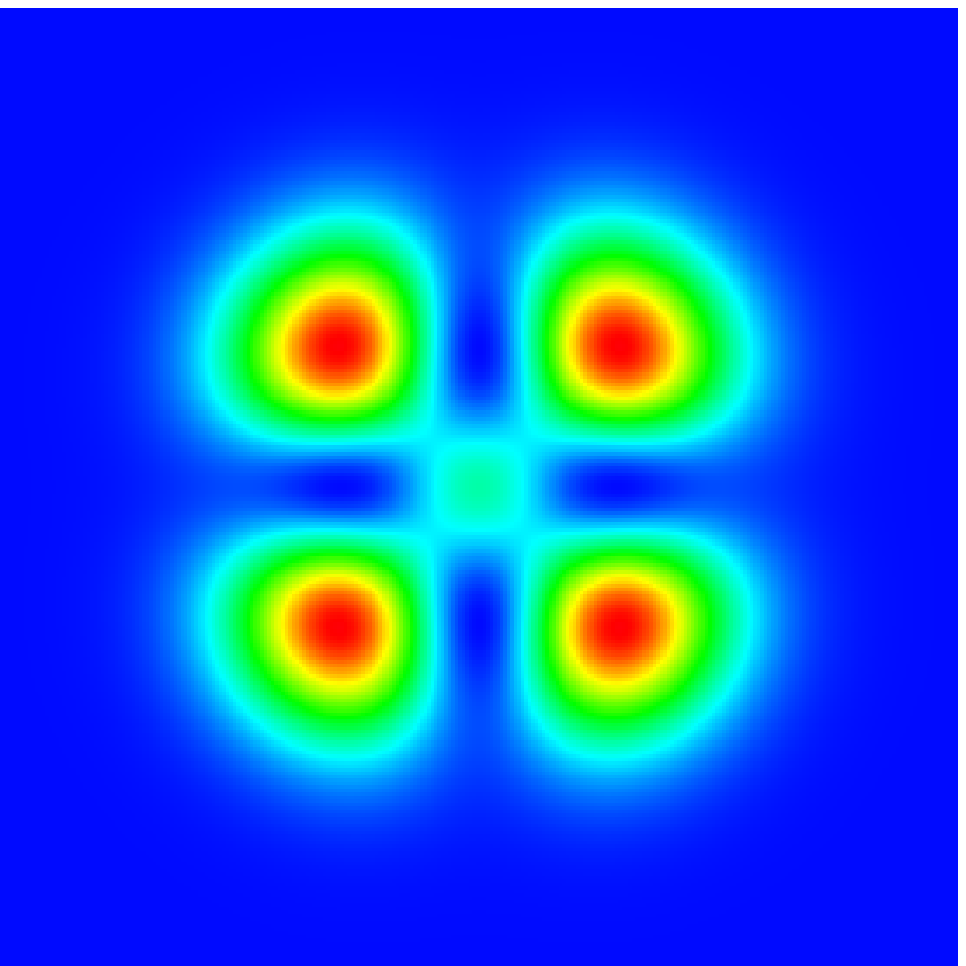}\\
\end{center}
\caption{\label{modes} Transverse intensity patterns in the
corotating frame of the $(1,1)$ mode of cavity I (top), which is
specified by $(g_{\mathrm{s}},g_{\mathrm{c}})
=(\frac{3}{4},\frac{1}{2})$ and destabilized by rotation, and
cavity II (bottom), which is specified by
$(g_{\mathrm{s}},g_{\mathrm{c}}) =(-\frac{3}{4},-\frac{1}{2})$ and
stabilized by rotation, for increasing rotation frequencies. From
left to right it increases from 0 to $\Omega_{0}/6$ for cavity I
and from $0.21\Omega_{0}$ to $\Omega_{0}/4$ for cavity II. The
plots show the mode patterns close to the spherical mirror and the
vertical direction corresponds to the direction in which the
cylindrical mirror is flat.}
\end{figure}

By using the expression of the ray matrix in the corotating frame
(\ref{mrt}) and the stability criterion that its eigenvalues must
have a unit length, we find the stable, partially stable and
unstable sections in the configuration space
$(g_{\mathrm{s}},g_{\mathrm{c}})$ for different values of the
rotation frequency. The results are shown in Fig. \ref{stab}.
These plots reveal that, already at relatively small rotation
frequencies, quite drastic changes take place. For instance, near
$(g_{\mathrm{s}},g_{\mathrm{c}})=(1,0)$ stable configurations are
destabilized to become (partially) unstable, while partially
stable geometries near the negative $g_{\mathrm{c}}$axis are
stabilized by the rotation. An optical cavity can thus both lose
and gain the ability to confine light due to the fact that it
rotates. It is noteworthy that some configurations, for example
those with small and positive $g_{\mathrm{s}}$ and
$g_{\mathrm{c}}$, are first partially destabilized by rotation, but
retrieve stability if the rotation frequency is further increased.
Another remarkable feature of the plots in Fig. \ref{stab} is the
absence of partially stable areas in the lower right plot. As we
will argue below, this is more generally true for the rotation
frequency $\Omega_{0}/4$. In this specific case, the boundaries of
stability are given by the hyperbolas $g_{c}=1/(2g_{s})$ and
$g_{c}=1/(2g_{s}-1)$ and their asymptotes.

As we have recently shown \cite{Habraken08}, the structure of the
modes of a rotating cavity is fully determined by the eigenvectors
$\mu_{i}$. The modes are defined as corotating solutions of the
time-dependent paraxial wave equation (\ref{tdpweu}) that vanish on
the mirror surfaces. Geometric stability comes in as the necessary
and sufficient requirement for them to exist. Here, we illustrate the
effect of rotational (de)stabilization on the mode structure by
considering two cases of a cavity with a spherical and a cylindrical
mirror. Cavity I is specified by
$(g_{\mathrm{s}},g_{\mathrm{c}})=(\frac{3}{4},\frac{1}{2})$. It is
stable in the absence of rotation and destabilized at a rotation
frequency $\Omega=\Omega_{0}/6$. Cavity II is specified by the
parameter values $(g_{\mathrm{s}},g_{\mathrm{c}})
=(-\frac{3}{4},-\frac{1}{2})$. It is partially stable in the
absence of rotation and stabilized by rotation at
$\Omega\simeq0.2098\Omega_{0}$. The effect of rotation on the
spatial structure of the modes of cavities I and II is shown in
Fig. \ref{modes}. The upper frames show the transverse spatial
structure on the spherical mirror of the $(1,1)$ mode of cavity I.
From left to right the rotation frequency increases from $0$ to
$0.166\Omega_{0}$ in equal steps. In the absence of rotation (left
frame) the mode is an astigmatic Hermite-Gaussian mode. Due to
rotation, the mode is deformed to a generalized Gaussian mode with
a nature in between Hermite-Gaussian and Laguerre-Gaussian modes
\cite{Abramochkin04}. As a result, phase singularities or
so-called optical vortices \cite{Soskin}, which are visible as
points with zero intensity, appear. For rotation frequencies close
to $\Omega_{0}/6$, the mode loses its confinement in the vertical
direction. This reflects the fact that the cavity approaches a
region of partial instability. The lower frames in Fig.
\ref{modes} show the intensity pattern on the spherical mirror of
the $(1,1)$ mode of cavity II, which is stabilized by rotation.
From left to right the rotation frequency is increased from
$0.21\Omega_{0}$ to $0.25\Omega_{0}$ in equal steps. As a result
of the rotation we retrieve a mode that is confined in both
directions and is similar to a Hermite-Gaussian mode. Deformation
of the mode due to the rotation is more pronounced for even larger
values of the rotation frequency.

Obviously, the horizontal and vertical directions in Fig.
\ref{modes}, which correspond to the curved and flat directions of
the cylindrical mirror, are lines of symmetry. In the special case
of a rotation frequency $\Omega_{0}/4$, the cylindrical mirror is
rotated over $\pi/2$ after each round trip so that its orientation
is periodic with two round-trip times as a period. This causes the
diagonal lines between the horizontal and vertical directions to
be lines of symmetry of the round-trip ray matrix (\ref{mrt}) and
the intensity patterns. This explains the apparent absence of
astigmatism in the lower right plot of Fig. \ref{modes}. This
additional symmetry also causes the four eigenvalues $\lambda_{i}$
to have the same absolute value, which explains the absence of
partial stability in the lower right plot of Fig. \ref{stab}.

\begin{figure}[!t]
\begin{center}
\includegraphics[width=4cm]{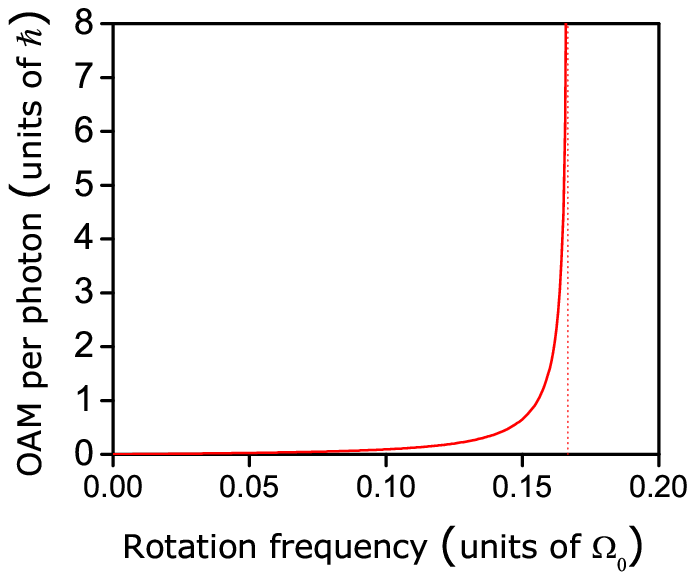}\hspace{0.3cm}
\includegraphics[width=4cm]{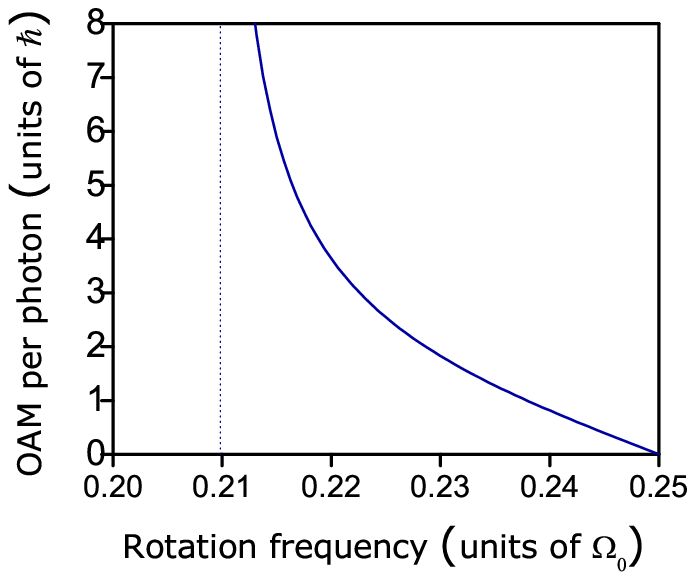}
\end{center}
\caption{\label{OAM} Dependence on the rotation frequency of the
orbital angular momentum per photon in the $(1,1)$ mode of cavity
I (left), which is destabilized by rotation, and cavity II (right),
which is stabilized by rotation.}
\end{figure}

Though the intensity patterns of the modes are aligned along the
axes of the cylindrical mirror, their phase patterns are not.
These attain a twist that is a signature of orbital angular
momentum \cite{Molina-Terriza07,Visser04}, proportional to $\int
d\rho v^{\ast}(\rho,z)\hat{L}_{\mathrm{z}}v(\rho,z)$. The dependence of this
orbital angular momentum in the $(1,1)$ mode of cavity I on the
rotation frequency is shown in Fig. \ref{OAM} (left plot). The
orbital angular momentum shows a divergence at $\Omega_{0}/6$,
which arises from the induced instability of the cavity. The
opposite happens for cavity II (right plot), which is stabilized
by rotation. In this case the orbital angular momentum decreases
with increasing rotation frequencies and eventually vanishes for
$\Omega=\Omega_{0}/4$ due to the additional symmetry at this
specific rotation frequency. The vanishing orbital angular
momentum does not imply that there is no vorticity in the modes at
this rotation frequency. The two contributions to the orbital
angular momentum add up to zero for modes with two equal mode
numbers.

In this paper, we have investigated rotationally induced
transitions between the areas of stability and partial instability
of an astigmatic two-mirror cavity. This is the first example of
an optical system where stability can be induced or removed by
rotation. Mechanical systems with dynamical stabilization are the
Paul trap and the gyroscope. The most obvious signatures of
rotational (de)stabilization are the modification of the mode
confinement and the divergence of the orbital angular momentum,
respectively shown in Figs. \ref{modes} and \ref{OAM}. The spatial
structure of these modes may be difficult to measure, but since
their orbital angular momentum components appear at different
frequencies due to the rotational Doppler shift
\cite{Nienhuis96,Courtial98}, it is possible to resolve the
divergence of the orbital angular momentum spectroscopically. The
effects of transverse rotations on the optical properties of a
cavity are significantly more complex than the resonance shifts
that are associated with small longitudinal displacements of the
mirrors. This may have important consequences in cavity-assisted
optomechanical experiments in which the rotational degrees of
freedom of a mirror are addressed.

Though the setup that we have studied here is rather
specific, our method, which is exact in the paraxial limit, can be
applied to more complex optical systems. Moreover, it should also
be applicable to other, mathematically similar, wave-mechanical
systems. Examples include the quantum-mechanical description of a
particle in a rotating, partially stable potential and rotating
acoustical cavities. In particular, the modification of the mode
confinement and the rotationally induced angular momentum are
expected to have analogues in such systems.

It is a pleasure to thank Eric~R.~Eliel and Bart-Jan Pors for
fruitful discussions and valuable suggestions regarding this
manuscript.

\end{document}